\documentclass[12pt]{iopart}

\begin{document}

\title{Exact models for isotropic matter}

\author{S Thirukkanesh\dag \ddag\  and S D
Maharaj\dag \footnote[3]{To whom correspondence should be addressed:
maharaj@ukzn.ac.za}}

\address{\dag\ Astrophysics and Cosmology Research Unit,
 School of Mathematical Sciences,
 University of KwaZulu-Natal,
 Durban 4041, South Africa
 }

\address{\ddag\ Present address: Department of
Mathematics, Eastern University, Chenkalady, Sri Lanka}

\begin{abstract}
We study the Einstein-Maxwell system of equations in spherically
symmetric gravitational fields for static interior spacetimes.
 The condition for pressure isotropy
  is reduced to a recurrence equation with variable, rational
coefficients. We demonstrate that this difference equation can be
solved in general using mathematical induction. Consequently we can
find an explicit exact solution to the Einstein-Maxwell field
equations. The metric functions, energy density, pressure and the
electric field intensity can be found explicitly. Our result
contains models found previously including the neutron star model of
Durgapal and Bannerji. By placing restrictions on  parameters
arising in the general series we show that the series terminate and
there exist two linearly independent solutions. Consequently it is
possible to find exact solutions in terms of elementary functions,
namely polynomials and algebraic functions.
\end{abstract}

\maketitle

\section{Introduction}
The first  spacetime studied, modelling the interior of the
relativistic sphere with uniform  energy density, is the
Schwarzschild interior solution which was found about ninety years
ago. Since then considerable time and energy has been spent in
finding exact solutions to the Einstein field equations
 for the interior spacetime that matches to the Schwarzschild exterior.
 The principal reason for  this activity is that  these
solutions to the field equations for spherically symmetric
gravitational fields  are necessary in the description of compact
objects in relativistic astrophysics. The models generated are used
to describe relativistic spheres with strong  gravitational fields
as is the case in neutron stars.  The detailed treatments of
Stephani {\it et al} \cite{Stephani} and Delgaty and Lake
\cite{Lake} for static, spherically symmetric models provide a
comprehensive collection of interior spacetimes, satisfying a
variety of criteria for physical admissability, that match to the
Schwarzschild exterior spacetime. It is important to note that only
a few of these solutions correspond to nonsingular gravitational
potentials with a physically acceptable energy momentum tensor and a
barotropic equation of state. A sample of the exact solutions to the
field equations, which satisfy all the physical requirements for a
relativistic star, is contained in the models of the Durgapal and
Bannerji \cite{Durgapal}, Durgapal and Fuloria \cite{Fuloria}, Finch
and Skea \cite{Finch}, Ivanov \cite{Ivanov}, Lake \cite{Lake2},
Maharaj and Leach \cite{Maharaj} and Sharma and Mukherjee
\cite{SharmaMukherjee2002}, amongst others.

In the past most of the solutions found have been obtained in an
{\it ad hoc} manner. We expect that a more systematic and formal
study, such as the treatment of Maharaj and Chaisi \cite{Chaisi} for
anisotropic matter, should lead to  new classes of solution. Clearly
there is a need to systematically study the mathematical properties
and features of the underlying nonlinear differential equations. The
analysis of John \cite{John} indicates that reducing the condition
of pressure isotropy to a recurrence relation with real, rational
coefficients leads to new mathematical and physical insights in the
Einstein equations for neutral matter. We attempt to perform similar
analysis here in the coupled Einstein-Maxwell equations for charged
matter. In this more general case we find that the condition of
pressure isotropy leads to a new recurrence relation which can be
solved in general. The Einstein-Maxwell system is important in the
description of a relativistic star in the presence of an
electromagnetic field. It is interesting to observe that, in the
presence of charge, the gravitational collapse of a spherically
symmetric distribution of matter to a point singularity may be
avoided. In this situation the gravitational attraction is
counterbalanced by the repulsive Coulombian force in addition to the
pressure gradient. Consequently the Einstein-Maxwell system, for a
charged star, has attracted considerable attention in various
physical investigations including Mukherjee \cite{Mukherjee2} and
Sharma {\it et al} \cite{Sailo}.

In this paper we seek new exact solutions to the Einstein field
equations, using a systematic series analysis, which may be used to
describe the interior gravitational profile of a relativistic
sphere. The approach produces a number of difference equations which
we demonstrate can be solved from first principles.
 In section 2 we first express the Einstein equations for neutral matter
 and the Einstein-Maxwell system for charged matter as
equivalent sets of differential equations utilising a transformation
due to Durgapal and Bannerji \cite{Durgapal}. We choose particular
forms for one of the gravitational potentials and the the electric
field intensity, which we believe has not been studied before. This
enables us to simplify the condition of pressure isotropy in section
3 to a second order linear equation in the remaining gravitational
potential. We assume a series form for this function which yields a
difference equation which we manage to solve using mathematical
induction. It is then possible to exhibit a new exact solution to
the Einstein-Maxwell field equations which can be written explicitly
as shown in section 4. We consider two particular cases in section 5
which contain exact solutions found previously.  In section 6 we
demonstrate that it is possible to find two linearly independent
solutions to the condition of pressure isotropy in terms of
elementary functions by placing restrictions on parameters,  that
appear in the general solution, thereby  permitting the series to
terminate. In section 7 we express the general solution of the
Einstein Maxwell system in terms of polynomials and algebraic
functions. We briefly discuss some of the physical properties of our
solutions in section 8.

\section{Field equations}
On physical grounds the gravitational field should be static and
spherically symmetric for describing the internal structure of a
dense compact relativistic sphere. Consequently we can find
coordinates $(t,r, \theta,\phi)$ such that the line element is of
the form
    \begin{equation}
    \rmd s^2 = -e^{2\nu(r)}\rmd t^2 + e^{2\lambda(r)} \rmd r^2
    + r^{2} (\rmd \theta^2 +\sin^{2}\theta \rmd \phi^{2}). \label{spherical}
    \end{equation}
For neutral perfect fluids the Einstein field equations can be
expressed as follows  \numparts
    \begin{eqnarray}
\frac{1}{r^2}  [r(1-e^{-2\lambda})]'& =& \rho \label{perfect-a}
\\   -\frac{1}{r^2}\ (1-e^{-2\lambda})+\frac{2\nu'}{r}\
e^{-2\lambda}& =& p \label{perfect-b}  \\   e^{-2\lambda}
\left(\nu''+{\nu}'^{2} +\frac{\nu'}{r}  -\nu' \lambda'-
\frac{\lambda'}{r}\right) & = & p \label{perfect-c}
\end{eqnarray}
    \endnumparts
for the spherically symmetric line element (\ref{spherical}).
 The energy density $\rho$ and the pressure $p$ are measured relative
to the comoving fluid 4-velocity $u^a = e^{-\nu} \delta^{a}_{\,0}$
and primes denote differentiation with respect to the radial
coordinate $r$. In the field equations
(\ref{perfect-a})-(\ref{perfect-c}) we are using units where the
coupling constant $\frac{8 \pi G}{c^4} = 1$ and the speed of light
$c=1$. The system of equations (\ref{perfect-a})-(\ref{perfect-c})
governs the behaviour
 of the gravitational field for a neutral perfect fluid.
 A different but equivalent form of the field
equations is obtained if we define a new independent variable $x$,
and new functions $y$ and $Z$, as follows
\begin{equation}  A^2 y^2 (x) = e^{2 \nu(r)}, \,\,\, Z(x) =
e^{-2 \lambda(r)}, \,\,\, x = C r^2 \label{transf}
 \end{equation}
 so that the line element (\ref{spherical}) becomes
 \[ \rmd s^2 = -A^2 y^2 \rmd t^2 + \frac{1}{4CxZ} \rmd x^2
    + \frac{x}{C}(\rmd \theta^2 +\sin^{2}\theta \rmd \phi^{2}).
 \]
 In the equations (\ref{transf}) the quantities $A$ and $C$ are
 arbitrary constants. Under
the transformation (\ref{transf}) the system
(\ref{perfect-a})-(\ref{perfect-c})  has the form \numparts
    \begin{eqnarray}
    \frac{1-Z}{x}-2\dot{Z} &=&\frac{\rho}{C}            \label{Durg-a}\\
    4Z\frac{\dot{y}}{y} + \frac{Z-1}{x} &=& \frac{p}{C} \label{Durg-b}\\
    4Zx^2 \ddot{y} + 2\dot{Z}x^{2} \dot{y} + (\dot{Z}x-Z+1)y&=&0    \label{Durg-c}
    \end{eqnarray} \endnumparts
where the dots denotes differentiation with respect to the variable
$x$. The set (\ref{Durg-a})-(\ref{Durg-c}) is a system of three
equations in the four unknowns $\rho, p, y$ and $Z$.

A generalisation of the system (\ref{perfect-a})-(\ref{perfect-c})
is the Einstein-Maxwell field equations  given by \numparts
\begin{eqnarray}
\frac{1}{r^{2}} \left[ r(1-e^{-2\lambda}) \right]' & = & \rho + \frac{1}{2}E^{2}
\label{perfectc-a} \\
 - \frac{1}{r^{2}} \left( 1-e^{-2\lambda} \right) +
\frac{2\nu'}{r}e^{-2\lambda} & = & p -\frac{1}{2}E^{2} \label{perfectc-b} \\
 e^{-2\lambda} \left( \nu'' + \nu'^{2} + \frac{\nu'}{r} -
\nu'\lambda' - \frac{\lambda'}{r} \right) & = & p + \frac{1}{2}E^{2}
\label{perfectc-c} \\
 \sigma & = & \frac{1}{r^{2}} e^{-\lambda}(r^{2}E)'. \label{perfectc-d}
\end{eqnarray} \endnumparts
where $E$ is the electric field intensity and $\sigma$ is the charge
density. When the electric field $E=0$ then the Einstein-Maxwell
equations (\ref{perfectc-a})-(\ref{perfectc-d}) reduce to the
Einstein equations ({\ref{perfect-a})-({\ref{perfect-c}) for neutral
matter. The system of equations
(\ref{perfectc-a})-(\ref{perfectc-d}) governs the behaviour
 of the gravitational field for a charged perfect fluid. If we
 use  the transformation (\ref{transf})
then the Einstein-Maxwell system
(\ref{perfectc-a})-(\ref{perfectc-d}) becomes \numparts
\begin{eqnarray} \frac{1-Z}{x} - 2\dot{Z} & = & \frac{\rho}{C} +
 \frac{E^{2}}{2C} \label{Durgc-a}\\
 4Z\frac{\dot{y}}{y} + \frac{Z-1}{x} & = & \frac{p}{C} -
 \frac{E^{2}}{2C}\label{cubic17-b} \label{Durgc-b}\\
 4Zx^{2}\ddot{y} + 2 \dot{Z}x^{2} \dot{y} + \left(\dot{Z}x -
Z + 1 - \frac{E^{2}x}{C}\right)y & = & 0 \label{Durgc-c}\\
 \frac{\sigma^{2}}{C} & = & \frac{4Z}{x} \left(x \dot{E} + E
\right)^{2} \label{Durgc-d}
\end{eqnarray} \endnumparts
which may be easier to integrate in certain situations.

\section{Specifying $Z$ and $E$}
We examine a particular form of the Einstein-Maxwell
 field equations (\ref{Durgc-a})-(\ref{Durgc-d}) by making explicit choices for the gravitational
potential $Z$ and the electric field intensity $E$. The system
(\ref{Durgc-a})-(\ref{Durgc-d})
 comprises four equations in six unknowns $Z, y, \rho, p, E$ and $\sigma$.
 By specifying the gravitational potential $Z$ and electric field
intensity $E$ we are in a position to integrate the condition of
pressure isotropy (\ref{Durgc-c}).
 The solution of the Einstein-Maxwell system then follow. We make the choice
\begin{equation}
Z = \frac{1+kx}{1+x} \label{Z}
\end{equation}
where $k$ is a real constant. In (\ref{Z}) we take $k\not= 1$. If
$k=1$ then the metric function $e^{2\lambda}$=1 and the energy
density is $ \rho= -\frac{E^{2}}{2}.$ To avoid negative energy
densities, which are not physical for barotropic stars,
 we consequently take $k\not=1$. The choice (\ref{Z}) was also made
 by Maharaj and Mkhwanazi \cite{MaharajM} and  in their
analysis of uncharged stars. Our objective is to confirm that
 this type of potential is also consistent with nonvanishing
electromagnetic fields. Note that our choice contains, as a special
case, the Durgapal and Bannerji \cite{Durgapal} solution, which is
widely applied as a relativistic model for neutral stars. Only the
solutions for the cases $k= \frac{1}{2}$ and $k=-\frac{1}{2}$ were
documented previously for the uncharged case when $E=0$. Other
physically reasonable choices of the gravitational potential $Z$ are
possible; we have chosen the form (\ref{Z}) as it produces charged
and uncharged solutions which are necessary
 for a realistic model.

  Upon substituting (\ref{Z}) in equation (\ref{Durgc-c}) we obtain
\begin{equation}
4(1+kx)(1+x) \ddot{y}+2(k-1) \dot{y}+
\left(1-k-\frac{E^{2}(1+x)^{2}}{Cx}\right)y = 0. \label{isotropy}
\end{equation}
 As the differential equation (\ref{isotropy}) is difficult to solve we first
introduce the transformation \numparts
\begin{eqnarray}
1+x & = & KX \label{9a}\\
K   & = & \frac{k-1}{k} \label{9b}\\
Y(X)& = & y(x) \label{9c}
\end{eqnarray} \endnumparts
 so as to obtain  a more convenient form.
Substituting  (\ref{9a})-(\ref{9c}) in the differential equation
(\ref{isotropy}) we obtain
\begin{equation}
4X(1-X)\frac{d^{2}Y}{dX^{2}} - 2\frac{dY}{dX} +
 \left(K + \frac{K^{2}(1-K)E^{2}X^{2}}{C(KX-1)} \right)Y = 0
 \label{isotropy2}
\end{equation}
in terms of the new dependent and independent variables $Y$ and $X$
respectively. The differential equation (\ref{isotropy2})
 may be integrated once  the electric field $E$ is specified.
A variety of choices for $E$ is possible but only a few are
physically reasonable and generate solutions in closed form.  We
observe that the particular choice
\begin{equation}
E^{2} = \frac{\alpha C}{K^{2}(1-K)} \frac{KX-1}{X^{2}}
\label{charge}
\end{equation}
where $\alpha$ is a constant, simplifies (\ref{isotropy2}). The
electric field defined in (\ref{charge}) vanishes at the centre of
the star, and remains continuous and bounded in the interior of the
star for a wide range of  values of the parameter $k$. Thus the
choice for $E$ is physically reasonable and it is a useful form to
study the gravitational behaviour of charged stars. With the help of
(\ref{charge}) we find that (\ref{isotropy2}) takes the simpler form
\begin{equation}
4X(1-X)\frac{d^{2}Y}{dX^{2}} - 2\frac{dY}{dX} + (K +\alpha)Y = 0.
\label{isotropy3}
\end{equation}
This is a special case of the hypergeometric equation. When
$\alpha=0$
 the differential equation (\ref{isotropy3}) becomes
\begin{equation}
4X(1-X)\frac{d^{2}Y}{dX^{2}} - 2\frac{dY}{dX} + K Y = 0
\label{isotropy4}
\end{equation}
and there is no charge.

\section{General series solution}
Since (\ref{isotropy3}) is the hypergeometric equation it is not
 possible to express the general solution in terms of
elementary functions for all $K + \alpha$. In general the solution
will be given in terms of special functions.
 The representation of the solution in a simple form is necessary
for a detailed physical analysis. Hence we attempt to obtain a
general solution to the differential equation
 (\ref{isotropy3}) in series form. Later we show that it is possible to extract
 solutions in terms
 of algebraic functions and polynomials.

 Since $X = 0$ is a regular singular point of the differential
equation (\ref{isotropy3}) we can apply the method of Frobenius
about $X=0$. We assume that the solution of the differential
equation (\ref{isotropy3}) is of the form
\begin{equation}
Y = \sum_{n=0}^{\infty}c_{n}X^{n+r}, \,\,\, c_{0}\not= 0
\label{series}
\end{equation}
where $c_{n}$ are the coefficients of the series and $r$ is a
constant. For a legitimate solution we need to determine the
coefficients $c_{n}$ as well as the parameter ${r}$.
 Substituting
(\ref{series}) in the differential equation (\ref{isotropy3}) we
obtain
\begin{eqnarray}
\fl 2c_{0}r[2(r-1)-1]X^{r-1} + & &  \nonumber \\
 \fl \sum_{n=0}^{\infty} \left( 2c_{n+1}(n+1+r)[2(n+r)-1]-
c_{n}[4(n+r)(n+r-1)-(K+\alpha)]\right) X^{n+r} & = & 0.
\label{series2}
\end{eqnarray}
The coefficients of the various powers of $X$ must vanish. Equating
the coefficient of $X^{r-1}$ in (\ref{series2}) to zero we obtain
the indicial equation
\[ 2c_{0}r[2(r-1)-1] = 0. \]
Since $c_{0}\not=0$ we must have $r=0$  or $r= \frac{3}{2}.$
 Equating the coefficient of $X^{n+r}$ in (\ref{series2}) to zero we obtain
\begin{equation}
c_{n+1}  =  \frac{4(n+r)(n+r-1)-(K+\alpha)}{2(n+1+r)
[2(n+r)-1]}c_{n} , \,\,\, n\geq 0  \label{difference}
\end{equation}
which is the fundamental difference equation governing the structure
of the solution.

 We can establish a general structure for the
coefficients by considering the leading terms. The coefficients
$c_1, c_2,c_3, \dots $can all be written in terms of the leading
coefficient $c_0$ and we generate the expression
\begin{equation}
c_{n+1} =
\prod_{p=0}^{n}\frac{4(p+r)(p+r-1)-(K+\alpha)}{2(p+1+r)[2(p+r)-1]}c_{0}
\label{difference2}
\end{equation}
where the symbol $\prod$ denotes multiplication. It is possible to
establish that the result (\ref{difference2})
 holds for all nonnegative integers using the principle of mathematical
 induction.

  We can now generate two linearly independent solutions, $y_1$ and
$y_2,$
 from (\ref{series}) and (\ref{difference2}). For the parameter value $r=0$
 we obtain the first solution
\begin{eqnarray}
Y_{1} &=& c_{0}
\left[1+\sum_{n=0}^{\infty}\prod_{p=0}^{n}\frac{4p(p-1)
-(K+\alpha)}{2(p+1)(2p-1)}X^{n+1} \right] \nonumber \\
 y_{1} &=& c_{0}
\left[1+\sum_{n=0}^{\infty}\prod_{p=0}^{n}\frac{4p(p-1)
-(K+\alpha)}{2(p+1)(2p-1)} \left(\frac{1+x}{K}\right)^{n+1} \right].
\label{y1}
\end{eqnarray}
For the parameter value $r=\frac{3}{2}$ we obtain the second
solution
\begin{eqnarray}
\fl Y_{2} &=& c_{0}X^{\frac{3}{2}} \left[1+\sum_{n=0}^{\infty}
\prod_{p=0}^{n}\frac{(2p+3)(2p+1)-(K+\alpha)}{(2p+5)(2p+2)}X^{n+1}
\right] \nonumber \\
 \fl y_{2} &=& c_{0}\left(\frac{1+x}{K}
\right)^{\frac{3}{2}}
\left[1+\sum_{n=0}^{\infty}\prod_{p=0}^{n}\frac{(2p+3)(2p+1)-
(K+\alpha)}{(2p+5)(2p+2)}\left(\frac{1+x}{K}\right)^{n+1} \right].
\label{y2}
\end{eqnarray}
Thus the general solution to the differential equation
(\ref{isotropy}),
 for the choice (\ref{charge}), is given by
\begin{equation}
y = a y_{1}(x) + by_{2}(x) \label{y}
\end{equation}
where $a$ and $b$ are arbitrary constants, $ K = \frac{k-1}{k}$ and
$y_{1}$ and $y_{2}$ are given by (\ref{y1}) and (\ref{y2})
respectively.
 By inspection it is clear that $y_1$ and $y_2$ are linearly independent functions.
From (\ref{Durgc-a})-(\ref{Durgc-d}), (\ref{y1}) and (\ref{y2}) the
general solution to the Einstein-Maxwell system becomes \numparts
\begin{eqnarray}
e^{2\lambda}   & = & \frac{1+x}{1+kx}  \label{solution1-a}\\
e^{2\nu}       & = & A^{2}y^{2} \label{solution1-b}\\
\frac{\rho}{C} & = & \frac{(1-k)(3+x)}{(1+x)^{2}}
- \frac{\alpha kx}{2(1+x)^{2}} \label{solution1-c} \\
\frac{p}{C}    & = & 4\frac{(1+kx)}{(1+x)}\frac{\dot{y}}{y}+
\frac{(k-1)}{(1+x)} + \frac{\alpha kx}{2(1+x)^{2}} \label{solution1-d}\\
\frac{E^{2}}{C}& = & \frac{\alpha kx}{(1+x)^{2}} \label{solution1-e}
\end{eqnarray} \endnumparts
where $y = ay_{1}(x) + by_{2}(x)$. We believe that
(\ref{solution1-a})-(\ref{solution1-e}) is a new solution  to the
Einstein-Maxwell field equations.

\section{Particular cases}
From the Einstein-Maxwell solution
(\ref{solution1-a})-(\ref{solution1-d}) we can generate a number of
physically reasonable charged and uncharged models for particular
choices of $k$ and $\alpha$. If we set $\alpha =0$ then } \numparts
\begin{eqnarray}
e^{2\lambda}   & = & \frac{1+x}{1+kx} \label{sol2-a}\\
e^{2\nu}       & = & A^{2}y^{2}\label{sol2-b} \\
\frac{\rho}{C} & = & \frac{(1-k)(3+x)}{(1+x)^{2}}  \label{sol2-c}\\
\frac{p}{C}    & = & 4\frac{(1+kx)}{(1+x)}\frac{\dot{y}}{y}+
 \frac{(k-1)}{(1+x)} \label{sol2-d}
\end{eqnarray} \endnumparts
which corresponds to a neutral relativistic star. We believe
 that the uncharged solution (\ref{sol2-a})-(\ref{sol2-d}) is also a new solution to
 the Einstein field equations (\ref{Durg-a})-(\ref{Durg-c}). It does not appear in
the comprehensive list of solutions presented by Delgaty and Lake
\cite{Lake}. In the solutions
(\ref{solution1-a})-(\ref{solution1-e}) and
(\ref{sol2-a})-(\ref{sol2-d}) the gravitational potentials $\lambda$
and $\nu$ are well behaved. Clearly the energy density $\rho$ is
positive at the origin if we choose $k < 1$. The pressure is finite
at the origin. These are desirable features in a stellar model.

When $K+\alpha =3$ the series in (\ref{y2}) terminates. It is then
possible to write the exact solution to the Einstein-Maxwell system
 in terms of elementary functions. The explicit form of the
solution is given by \numparts
\begin{eqnarray}
\fl e^{2\lambda}   & = & \frac{(K-1)(1+x)}{(K-1-x)} \label{sol3-a}\\
\fl e^{2\nu}       & = & A^{2}\left[c_{1}(1+x)^{\frac{3}{2}}
+ c_{2}(K-1-x)^{\frac{1}{2}}(K+2+2x) \right]^{2} \label{sol3-b}\\
\fl \frac{\rho}{C} & = & \frac{2K(3+x)+ (3-K)x}{2(K-1)(1+x)^{2}} \label{sol3-c}\\
\fl \frac{p}{C}    & = & \frac{1}{K-1} \times \nonumber \\
 \fl              &   &   \frac{c_{1}(1+x)^{\frac{1}{2}}[5K-6-(K+6)x]
+
c_{2}(K-1-x)^{\frac{1}{2}}[4K-12-K^{2}-(12+2K)x]}{c_{1}(1+x)^{\frac{5}{2}}
 + c_{2}(1+x)(K-1-x)^{\frac{1}{2}}(K+2+2x)}  \nonumber \\
   \fl            &   & +\frac{(3-K)x}{2(1-K)(1+x)^{2}} \label{sol3-d}\\
\fl \frac{E^{2}}{C}& = & \frac{(3-K)x}{(1-K)(1+x)^{2}}
\label{sol3-e}
\end{eqnarray} \endnumparts
where $ K = (k-1)/k$ and $x=Cr^{2}$. Clearly
(\ref{sol3-a})-(\ref{sol3-e}) is a special case of the general
solution (\ref{solution1-a})-(\ref{solution1-e}). The solution
(\ref{sol3-a})-(\ref{sol3-e}) has the advantage of being given
completely in terms of elementary functions which makes an analysis
of the physical features of the model possible. When $K=3$ (i.e.
$k=-\frac{1}{2}$) and $\alpha=0$ we obtain \numparts
\begin{eqnarray}
e^{2\lambda}   & = & \frac{2(1+x)}{(2-x)} \label{sol4-a}\\
e^{2\nu}       & = & A^{2}\left[c_{1}(1+x)^{\frac{3}{2}}
+ c_{2}(2-x)^{\frac{1}{2}}(5+2x) \right]^{2} \label{sol4-b}\\
\frac{\rho}{C} & = & \frac{3(3+x)}{2(1+x)^{2}} \label{sol4-c}\\
\frac{p}{C}    & = & \frac{9}{2} \left[
\frac{c_{1}(1+x)^{\frac{1}{2}}(1-x) -
c_{2}(2-x)^{\frac{1}{2}}(1+2x)}{c_{1}(1+x)^{\frac{5}{2}} +
c_{2}(1+x)(2-x)^{\frac{1}{2}}(5+2x)} \right] \label{sol4-d}
\end{eqnarray} \endnumparts
for an uncharged relativistic stellar model. The special case
(\ref{sol4-a})-(\ref{sol4-d}) is the same as the result of Durgapal
and Bannerji \cite{Durgapal} and Maharaj and Mhkwanazi
\cite{MaharajM}. We point out that Maharaj and Mhkwanazi
\cite{MaharajM} had a numerical mistake in their calculation of the
pressure $p$ which has been corrected
 in our solution. We believe that the solution
 (\ref{sol3-a})-(\ref{sol3-d}) is important
in the study of charged stars as it contains the Durgapal and
 Bannerji \cite{Durgapal} model which has been shown to be consistent
with a realistic dense star.  Extensive studies of the Durgapal
 and Bannerji  solution, as indicated in the compendium by
 Delgaty and Lake \cite{Lake}, has proved
that all the criteria for physical acceptability
 are satisfied. It is consequently used in many astrophysical
 studies that model neutron stars.

\section{Terminating series}
The general solution (\ref{y}) can be expressed in terms of the
special functions, namely hypergeometric functions. For
 particular values of $K$ and $\alpha$ the series solution
 can be given in terms of elementary functions as demonstrated in section 5.
 This is possible in general because the series (\ref{y1}) and (\ref{y2}) terminate
for restricted values of the parameters $K$ and $\alpha$. Using this
feature we obtain two sets of general solutions in terms of
elementary functions, by determining the specific restriction on
$K+\alpha$ for a terminating series, as demonstrated in the
following sections.

\subsection{The first solution}
 On substituting $r=0$ in equation
(\ref{difference}) we obtain
\begin{equation}
c_{i+1}  =  \frac{4i(i-1)-(K+\alpha)}{(2i+2)(2i-1)}c_{i} ,
 \,\,\, i\geq 0. \label{difference3}
\end{equation}
If we set $K+\alpha =4n(n-1)$,  where $n$ is a fixed integer, then
$c_{n+1}=0$.  Clearly the subsequent coefficients $c_{n+2}, c_{n+3},
c_{n+4}, \dots$ vanish and equation (\ref{difference3}) has the
solution
\begin{equation}
    c_i    =  4n(n-1)\frac{(-4)^{i-1}(2i-1)(n+i-2)!}
{(2i)!(n-i)!}c_{0} , \,\,\, 1\leq i \leq n. \label{difference4}
\end{equation}
 Then from the
equations (\ref{series}) (when $r=0$) and (\ref{difference4}) we
obtain
\begin{equation}
Y_{1} = c_{0} \left[1+ 4n(n-1)\sum_{i=1}^{n}
\frac{(-4)^{i-1}(2i-1)(n+i-2)!}{(2i)!(n-i)!}X^{i} \right]
\label{solution5}
\end{equation}
where \( K+\alpha=4n(n-1). \)

On substituting $r = \frac{3}{2}$ in (\ref{difference}) we obtain
\begin{equation}
c_{i+1}  =  \frac{(2i+3)(2i+1)-(K+\alpha)} {(2i+5)(2i+2)}c_{i} ,
\,\,\, i\geq 0. \label{difference45}
\end{equation}
If we set $K+\alpha =(2n+3)(2n+1)$, where $n$ is a fixed integer
then $c_{n+1}=0$. Also the subsequent coefficients $c_{n+2},c_{n+3},
c_{n+4}, \dots$ vanish and  equation (\ref{difference45}) can be
solved to yield
 \begin{equation}
 c_i  =  \frac{3 (-4)^{i}(2i+2)(n+i+1)!}
 {(n+1)(n-i)!(2i+3)!}c_{0}, \,\,\, 1 \leq i \leq n.
\label{difference46}
\end{equation}
 Then from the equations (\ref{series}) (when $ r=\frac{3}{2}$ ) and
(\ref{difference46}) we obtain
\begin{equation}
Y_{1} = c_{0}X^{\frac{3}{2}} \left[ 1 + \frac{3}{(n+1)}
\sum_{i=1}^{n} \frac{(-4)^{i}(2i+2)(n+i+1)!}{(n-i)!(2i+3)!}X^{i}
\right] \label{solution6}
\end{equation}
where $K+\alpha = (2n+3)(2n+1)$. The  polynomial and algebraic
functions (\ref{solution5}) and (\ref{solution6}) comprise the first
solution of the differential equation (\ref{isotropy3}) for
appropriate values of $K+\alpha$.

\subsection{The second solution}
 We can use the form of  the particular solution in section 5, expressed in terms of
 elementary functions, to simplify the representation of the second
 solution.
The special solution  in section 5 contains terms  of the form  \(
(1-X)^{\frac{1}{2}}(1+2X) \) which is product of \(
(1-X)^{\frac{1}{2}}\) and a polynomial. This suggests that the
second solution in general is of the form
\[Y_{2} =(1-X)^{\frac{1}{2}}u(X) \]
where $ u(X)$ is an arbitrary function. We now take $Y_2$ to be the
generic second solution of (\ref{isotropy3}) and explicitly
determine $u(X)$. Equation (\ref{isotropy3}) gives
\begin{equation}
4X(1-X) \ddot{u} - 2(1+2X) \dot{u} + (1+K+\alpha)u =0
\label{isotropy5}
\end{equation}
where dots denote differentiation with respect to $X$.

 Since $X=0$
is a regular singular point of the differential equation
(\ref{isotropy5}) we can apply the method of Frobenius. We assume
that the solution is of the form
\begin{equation}
u = \sum_{n=0}^{\infty}c_{n}X^{n+r}, \,\,\, c_{0} \not= 0.
\label{series3}
\end{equation}
On substituting (\ref{series3}) in the differential equation
(\ref{isotropy5})
 we obtain
\begin{eqnarray}
\fl 2c_{0}r[2(r-1)-1]X^{r-1} + &   & \nonumber \\
\fl \sum_{n=0}^{\infty} \left(2c_{n+1}(n+1+r)[2(n+r)-1] -
c_{n}[4(n+r)^{2} - (1+K+\alpha)]\right)  X^{n+r} & = & 0.
\label{series4}
\end{eqnarray}
The coefficients of the various powers of $X$ have to vanish.
 Setting the coefficient of $X^{r-1}$ in (\ref{series4}) to zero we obtain
 the indicial equation
\[ 2c_{0}r[2(r-1)-1] = 0. \]
Since $c_{0}\not= 0$ we must have $r=0$ or $r= \frac{3}{2}$ as in
section 4. Equating the coefficient of $X^{n+r}$ in (\ref{series4})
to zero we obtain
\begin{equation}
c_{n+1}  =  \frac{4(n+r)^{2} -(1+K+\alpha)}{2(n+r+1)[2(n+r)-1]}
c_{n} \label{difference5}
\end{equation}
which is the basic difference equation governing the structure of
the solution.

We establish a general structure for the coefficients by considering
the leading terms. On substituting $r=0$ in equation
(\ref{difference5}) we obtain
\begin{equation}
c_{i+1}  =  \frac{4i^{2} -(1+K+\alpha)}{(2i+2)(2i-1)} c_{i}.
\label{difference6}
\end{equation}
We assume that \(K + \alpha = (2n+3)(2n+1)\) where $n$ is a fixed
integer. Then $c_{n+2}=0$ from (\ref{difference6}). Consequently the
remaining coefficients  $c_{n+3}, c_{n+4}, c_{n+5},\dots $ vanish
and equation (\ref{difference6}) has the solution
 \begin{equation}
 c_i  =  4(n+1)\frac{(-4)^{i-1}(2i-1)(n+i)!}{(2i)!(n-i+1)!}c_{0} ,
\,\,\, 1 \leq i \leq n+1. \label{difference77}
\end{equation}
 Then from equation
(\ref{series3})(when $r = 0$) and (\ref{difference77}) we obtain
\[ u = c_{0} \left[ 1+   4(n+1) \sum_{i=1}^{n+1}
\frac{(-4)^{i-1}(2i-1)(n+i)!}{(2i)!(n-i+1)!}X^{i} \right]. \] Hence
we have the result
\begin{equation}
Y_{2} = c_{0}(1-X)^{\frac{1}{2}} \left[ 1+  4(n+1)
\sum_{i=1}^{n+1}\frac{(-4)^{i-1}(2i-1)(n+i)!}{(2i)!(n-i+1)!}X^{i}\right]
\label{solution7}
\end{equation}
where $K + \alpha =(2n+3)(2n+1).$

 On substituting \( r= \frac{3}{2}\) in equation (\ref{difference5}) we obtain
\begin{equation}
c_{i+1}= \frac{(2i+3)^{2}-(1+K+\alpha)}{(2i+5)(2i+2)}c_{i}.
\label{difference7}
\end{equation}
We assume that \(K+\alpha = 4n(n-1)\) where $n$ is a fixed integer.
Then $c_{n-1}=0$ from (\ref{difference7}).   Consequently the
remaining coefficients $c_{n},c_{n+1}, c_{n+2}, \dots$ vanish and
 (\ref{difference7}) can be solved to yield
 \begin{equation}
 c_i  = \frac{3(-4)^{i}(2i+2)(n+i)!}{n(n-1)(2i+3)!(n-i-2)!}c_{0}, \,\,\, i \leq
 n-2. \label{difference8}
\end{equation}
Then from the equations (\ref{series3}) (when $r= \frac{3}{2}$) and
(\ref{difference8}) we obtain
\[u = c_{0} X^{\frac{3}{2}} \left[ 1+ \frac{3}{n(n-1)} \sum_{i=1}^{n-2}
\frac{(-4)^{i}(2i+2)(n+i)!}{(2i+3)!(n-i-2)!}X^{i} \right]. \] Hence
we have the result
\begin{equation}
Y_{2} = c_{0}(1-X)^{\frac{1}{2}} X^{\frac{3}{2}} \left[ 1+
\frac{3}{n(n-1)} \sum_{i=1}^{n-2}
\frac{(-4)^{i}(2i+2)(n+i)!}{(2i+3)!(n-i-2)!}X^{i} \right]
\label{solution8}
\end{equation}
where $K + \alpha = 4n(n-1).$

 The solutions (\ref{solution7}) and (\ref{solution8})
generate the second solution of the differential equation
(\ref{isotropy3}) which are clearly independent
 from the solutions (\ref{solution5}) and (\ref{solution6}).
The quantities (\ref{solution7}) and (\ref{solution8}) are products
of polynomials and algebraic functions.

\section{Elementary functions}

 Thus we have generated
general solutions to the differential equation (\ref{isotropy3}) by
restricting the values of $K+\alpha$ so that polynomials and product
of polynomials
 with algebraic functions are possible as solutions.  Collecting these
results we have the first category of solutions
\begin{eqnarray}
Y & = & a (1-X)^{\frac{1}{2}} \left[ 1+ 4(n+1) \sum_{i=1}^{n+1}
\frac{(-4)^{i-1}(2i-1)(n+i)!}{(2i)!(n-i+1)!}X^{i}\right]    \nonumber \\
  &   & + b X^{\frac{3}{2}} \left[ 1 + \frac{3}{(n+1)} \sum_{i=1}^{n}
\frac{(-4)^{i}(2i+2)(n+i+1)!}{(n-i)!(2i+3)!}X^{i} \right]
\label{solution9}
\end{eqnarray}
for \( K +\alpha = (2n+3)(2n+1)\) where $a$ and $b$ are arbitrary
constants. In terms of $x$ the solution (\ref{solution9}) becomes
\begin{eqnarray}
\fl y & = & a \left(\frac{K-1-x}{K} \right)^{\frac{1}{2}} \left[ 1+
4(n+1)
 \sum_{i=1}^{n+1}\frac{(-4)^{i-1}(2i-1)(n+i)!}{(2i)!(n-i+1)!}
\left(\frac{1+x}{K} \right)^{i}\right]    \nonumber \\
\fl  &   & + b \left(\frac{1+x}{K} \right)^{\frac{3}{2}} \left[ 1 +
\frac{3}{(n+1)} \sum_{i=1}^{n}
\frac{(-4)^{i}(2i+2)(n+i+1)!}{(n-i)!(2i+3)!}
\left(\frac{1+x}{K}\right)^{i} \right]. \label{solution10}
\end{eqnarray}
The second category of solutions is given by
\begin{eqnarray}
Y & = & a (1-X)^{\frac{1}{2}} X^{\frac{3}{2}} \left[ 1+
\frac{3}{n(n-1)}
 \sum_{i=1}^{n-2} \frac{(-4)^{i}(2i+2)(n+i)!}{(2i+3)!(n-i-2)!}X^{i} \right] \nonumber \\
&   & + b \left[1+ 4n(n-1)\sum_{i=1}^{n}
\frac{(-4)^{i-1}(2i-1)(n+i-2)!}{(2i)!(n-i)!}X^{i} \right]
\label{solution11}
\end{eqnarray}
for  \( K+ \alpha = 4n(n-1)\) where $a$ and $b$ are arbitrary
constants. In terms of $x$ the solution (\ref{solution11}) becomes
\begin{eqnarray}
\fl y & = & a \left(\frac{K-1-x}{K} \right)^{\frac{1}{2}}
 \left(\frac{1+x}{K} \right)^{\frac{3}{2}} \left[ 1+
\frac{3}{n(n-1)} \sum_{i=1}^{n-2}
\frac{(-4)^{i}(2i+2)(n+i)!}{(2i+3)!(n-i-2)!}
 \left(\frac{1+x}{K} \right)^{i} \right] \nonumber \\
 \fl &   & + b \left[1+ 4n(n-1)\sum_{i=1}^{n}
\frac{(-4)^{i-1}(2i-1)(n+i-2)!}{(2i)!(n-i)!} \left(\frac{1+x}{K}
\right)^{i} \right]. \label{solution12}
\end{eqnarray}

 It is remarkable that the solutions (\ref{solution10}) and (\ref{solution12}) are
expressed completely as combinations of polynomial
 and algebraic functions. It is rare to find general solutions,
 considering the nonlinearity of the gravitational interactions,
 to the field equations in terms of elementary functions.
  We have expressed our solutions in the
simplest possible form. This has the advantage of simplifying the
analysis of the physical properties of the dense star. Observe that
our treatment has combined both the charged and uncharged cases for
a relativistic star. If we substitute $\alpha=0$ in
(\ref{solution10}) and  (\ref{solution12}) then we can obtain the
solutions for the
 uncharged case directly. Consequently our approach has the unexpected
 but very desirable  feature
 of producing an uncharged (possibly new) solution to equations
 (\ref{Durg-a})-(\ref{Durg-c})
 from the charged solutions when $E=0$. We believe that the solutions obtained
in this paper to the Einstein (\ref{Durg-a})-(\ref{Durg-c}) and
Einstein Maxwell (\ref{Durgc-a})-(\ref{Durgc-d}) field  equations
have not been found before.

 From our general
class of solutions (\ref{solution10}) and (\ref{solution12}) it is
possible to generate particular  solutions found previously. If we
take $K=3$ and $\alpha =0$ $(n=0)$ then it is easy to verify that
the equation (\ref{solution10}) becomes
\[y =c_{1}(2-x)^{\frac{1}{2}}(5+2x)+ c_{2}(1+x)^{\frac{3}{2}}\]
where $c_{1}= a/9$ and $c_{2}=b/3^{\frac{3}{2}}$ are new arbitrary
constants. Thus we have regained the Durgapal and Bannerji
\cite{Durgapal} neutron star model.  Other explicit
 functional forms for $y$ are obtainable which could be useful in applications for
a dense star. As an example suppose that $K +\alpha=8 (n=2)$ then
from (\ref{solution12}) we obtain
\[y=c_1 (K-1-x)^{\frac{1}{2}} (1+x)^{\frac{3}{2}} +c_2
 \left[K^2 +4K(1+x)-8(1+x)^2 \right].\]
where $c_1 =a/K^2$ and $c_2 =b/K^2$ are new arbitrary constants. It
is now possible to generate an exact solution to the
Einstein-Maxwell system (\ref{Durgc-a})-(\ref{Durgc-d}) in terms of
elementary functions when $K+\alpha=8$. This is given by \numparts
\begin{eqnarray}
\fl e^{2\lambda}   & = & \frac{(K-1)(1+x)}{(K-1-x)} \label{sol13-a}\\
\fl e^{2\nu}       & = & A^{2}\left[c_1 (K-1-x)^{\frac{1}{2}}
(1+x)^{\frac{3}{2}}
 +c_2 \left(K^2 +4K(1+x)-8(1+x)^2 \right) \right]^{2} \label{sol13-b}\\
\fl \frac{\rho}{C} & = & \frac{2K(3+x)+ (K-8)x}{2(K-1)(1+x)^{2}} \label{sol13-c}\\
\fl \frac{p}{C}    & = & \frac{2 (K-1-x)^{\frac{1}{2}} \left[c_1
(1+x)^{\frac{1}{2}}(3K-4(1+x)) +8c_2
(K-1-x)^{\frac{1}{2}}(K-4(1+x))\right]}{(K-1)(1+x)
 \left[c_1 (K-1-x)^{\frac{1}{2}} (1+x)^{\frac{3}{2}} +c_2 \left(K^2
 +4K(1+x)-8(1+x)^2 \right) \right]} \nonumber\\
 \fl               &   & - \frac{K}{(K-1)(1+x)}+\frac{(K-8)x}{2(1-K)(1+x)^{2}}
 \label{sol13-d}\\
\fl \frac{E^{2}}{C}& = & \frac{(K-8)x}{(1-K)(1+x)^{2}}
\label{sol13-e}
\end{eqnarray} \endnumparts
where $ K = (k-1)/k$ and $x=Cr^{2}$.  The solution
(\ref{sol13-a})-(\ref{sol13-e}) is given in a simple form which
facilitates a physical analysis.

\section{Discussion}

We have found new solutions (\ref{solution1-a})-(\ref{solution1-e})
to the Einstein-Maxwell system using a systematic series analysis
that produces a number of difference equations which can be solved
in general. A useful feature of the approach is that we regain the
Durgapal and Bannerji neutron star model \cite{Durgapal} as a
special case which suggests that our class of solutions are
physically reasonable. We briefly consider some physical features of
the solutions of interest.

Firstly, in the general solution
(\ref{solution1-a})-(\ref{solution1-e}), when studying models of
charges spheres, we should consider only those values of $k$ for
which the energy density $\rho$, the pressure $p$ and the electric
field intensity $E$ are positive. The choice of $k$ must ensure that
the gravitational potential $e^{2\lambda}$ remains positive; the
remaining potential $e^{2\nu}$ is necessarily positive. Clearly a
wide range of charged spheres, with nonsingular potentials and
matter variables, are possible for relevant choices of $k$. The
interior metric (\ref{spherical}) must match to the
Reissner-Nordstrom exterior spacetime
\[
\fl \rmd s^2=-\left( 1-\frac{2M}{r} + \frac{Q^2}{r^2}\right)\rmd t^2
+
 \left( 1-\frac{2M}{r} + \frac{Q^2}{r^2}\right)^{-1} \rmd r^2 +
r^2(\rmd \theta^2 + \sin^2 \theta \rmd \phi^2)
\]
across the boundary $r=R$. This yields the relationships
\begin{eqnarray*}
 1-\frac{2M}{R} + \frac{Q^2}{R^2} &=&
 A^2 \left[ ay_1(CR^2) +by_2 (CR^2) \right]^2 \\
 \left( 1-\frac{2M}{R} + \frac{Q^2}{R^2}\right)^{-1} &=&
 \frac{1+CR^2}{1+kCR^2}
\end{eqnarray*}
between the constants $a,b,k, A$ and $C$. This shows that continuity
of the metric coefficients  across the boundary of the star is
easily achieved. The matching conditions at the boundary may place
restrictions on the function $\nu$ and its first derivative (for
uncharged matter) and the pressure may be nonzero (if there is a
surface layer of charge); there are sufficient free parameters
available to satisfy the necessary conditions that may arise from a
particular physical model under consideration.

Secondly, we observe that our solutions may be interpreted as models
for anisotropic spheres (which may be charged or uncharged) where
the parameter $\alpha$ plays the role of the anisotropy factor. The
solutions found depend smoothly on the parameter $\alpha$; isotropic
and uncharged solutions can be regained for $\alpha =0$. For recent
analyses of the physics of anisotropic matter see Chaisi and Maharaj
\cite{CM}, Dev and Gleiser \cite{Dev1}, \cite{Dev2} and Maharaj and
Chaisi \cite{Chaisi}.

\ack ST thanks South Eastern University for study leave and the
University of KwaZulu-Natal for a scholarship. SDM and ST thank the
National Research Foundation of South Africa for financial support.
We are grateful to the referees for insightful comments.

\end{document}